# Design Consideration on a Polarized Gas Target for the LHC

**Erhard Steffens**[1]
*FAU Erlangen-Nürnberg*
*Erlangen, Germany*
E-mail: Erhard.Steffens@fau.de

**Vittorio Carassiti**
*INFN Ferrara*
*Ferrara, Italy*
E-mail: vito@fe.infn.it

**Giuseppe Ciullo**
*Dipart. di Fisica, Univ. di Ferrara and INFN*
*Ferrara, Italy*
E-mail: ciullo@fe.infn.it

**Pasquale Di Nezza**
*INFN LNF*
*Frascati, Italy*
E-mail: Pasquale.Di.Nezza@cern.ch

**Paolo Lenisa**
*Dipart. di Fisica, Univ. di Ferrara and INFN*
*Ferrara, Italy*
E-mail: lenisa@fe.infn.it

**Luciano L. Pappalardo**
*Dipart. di Fisica, Univ. di Ferrara and INFN*
*Ferrara, Italy*
E-mail: pappalardo@fe.infn.it

**Alexander Vasilyev**
*PNPI Gatchina*
*Petersburg, Russia*
E-mail: Vasilyev_aa@pnpi.nrcki.ru

**Abstract:** Since 2017, the LHCSpin study group is investigating the installation of a HERMES-type polarized gas target (PGT) in front of the LHCb detector in order to perform Single-Spin Transverse Asymmetry (SSTA) measurements. In cooperation with LHC experts, the conditions for applying a PGT are being studied. As a viable option, a cold openable storage cell is considered. A key role for avoiding instabilities of the 7 TeV proton beam is the choice of a proper coating and the suppression of wake fields. A first warm ($\approx 300$ K) test storage cell is planned for installation in 2019 inside the VELO vessel, subject to final approval. It will improve the ongoing SMOG program of LHCb fixed target measurements, and will provide valuable experience of running a storage cell in the harsh LHC environment. The status of the design considerations on a PGT in the LHC beam and of the discussion of critical machine issues is presented.

---

[1]Speaker, corresponding author

## 1. Introduction

The LHCb Spectrometer at the LHC (CERN) detects particles from collisions at IP8. Primary and secondary vertices are reconstructed by means of the VErtex LOcator (VELO). The luminosity of the colliding bunches is measured via p-e scattering on the residual gas. In order to increase the precision of the luminosity measurements, a gas injection system (SMOG) for Ne (He, Ar) has been added. Starting in 2015, SMOG has been employed for Fixed-Target (FT) measurements, e.g. that of antiproton production in p-He FT collisions [1].

At present, the LHCSpin group works on the design to improve the SMOG FT measurements by adding a storage cell within the VELO vessel, upstream of the vertex detector. This project is called 'SMOG2'. A design has been presented to the LHCb collaboration in Fall 2018 (installation in 2019).

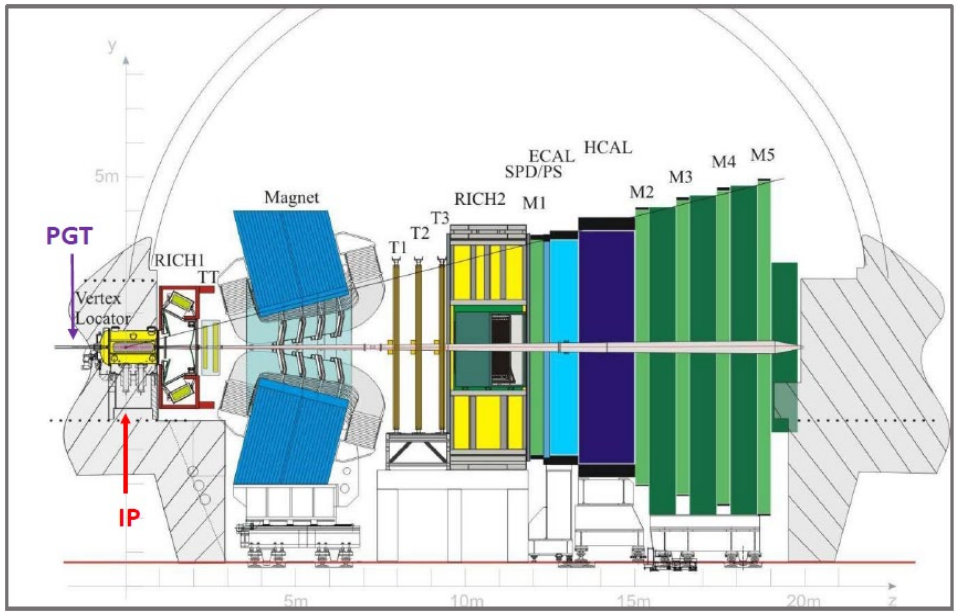

**Figure 1:** Sideview of the LHCb detector. Left: IP8 with VELO and the possible location of the target (PGT).

## 2. SMOG-upgrade SMOG2

The SMOG2 openable target cell (length L = 20 cm, inner diameter D = 1.0 cm) consists of two halves, connected to the two detector boxes (see Fig. 2). Prior to beam injection the cell is opened with the VELO boxes, and closed during stable operation of the LHC beam, when the VELO detector moves in. The gas is directly injected into the cell center resulting in a triangular pressure bump of length L and central density $\rho_0$ in the center [2]. The target areal density $\theta = \rho_0 \cdot L/2$ will be up to two orders of magnitude higher than for SMOG at the same gas flow rate (see Tab. 1). The cell is complemented by openable conducting surfaces, the Wake Field Suppressors (WFS), with smooth variation of the cross section in order to avoid excitation of wake fields. The application of SMOG for FT measurements is described in a recent paper [1] on antiproton production in p-He at $\sqrt{s_{NN}}$ = 110 GeV. Here, Helium has been injected by the SMOG system. From the reconstructed gas density and assuming a pumping speed on the VELO vessel of S = 500 l/s, a He flow rate $Q_0 = 1.3 \cdot 10^{-4}$ mbar l/s can be deduced. Data were analyzed from a fiducial region 0.8 m long, resulting in $\theta_{SMOG} = 5.6 \cdot 10^{11}$/cm$^2$ at this flow rate $Q_0$.

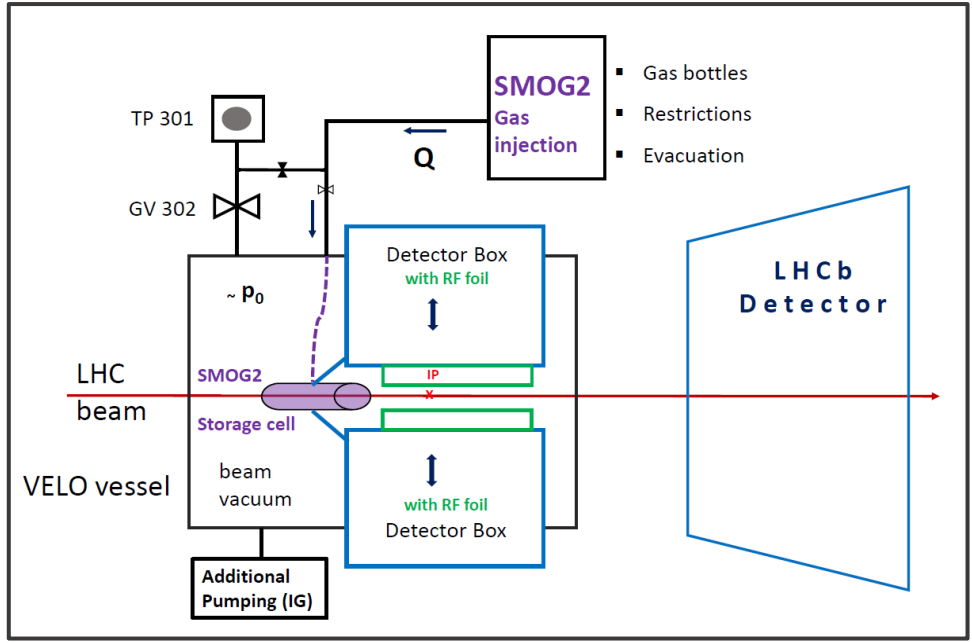

**Figure 2:** Principle of the SMOG2 target (see text). A gas feed system is employed to inject well-defined flow rates which allow for the precise prediction of the target density.

**2.1 Available densities with SMOG2 and comparison to SMOG**

The following **Table 1** shows the expected SMOG2 target densities for different gases at cell temperature 300 K, and for a typical SMOG flow rate $Q_0 = 1.3 \cdot 10^{-4}$ mbar l/s:

| Gas | $H_2$ | He | Ne | Ar | Kr |
|---|---|---|---|---|---|
| Areal density θ in $10^{12}$/cm$^2$ | 4.3 | 6.0 | 13.4 | 19.0 | 27.4 |
| Scale factor $\theta_{SMOG2} : \theta_{SMOG}$ | 7.6 | 10.7 | 24 | 34 | 49 |

The scale factor shows the improvement by the storage cell. The strong dependence on the molecular mass M comes from the $\sqrt{(M/T)}$ factor of θ, leading to a higher density for the heavier gases.

### 3. Design of the SMOG2 target cell system

The storage cell system [3] (see Fig. 3, left to right) is composed of the following elements: (i) upstream beam tube, (ii) cylindrical slotted WFS (yellow), (iii) conical WFS and cell tube with wings (blue), (iv) thin contact piece between cell and RF foil (light grey). Elements (ii) – (iv) consist of two halves and open with the detector boxes (dark grey). The cell system is made of Al (99.5 %) for the rigid parts (iii), and of Cu-Be for the flexible parts (ii), (iv). In the cell center, a small capillary for gas injection is visible. In the open position of the cell system, a free space of ≥ 50 mm in diameter is provided during injection and tuning of the beam. The cell is suspended by curved light-weight bars connected to the detector boxes, and by the contact pieces (iv) which fix the cell in the lateral degree of freedom. One cell halve is rigidly connected to the box, while the opposite 'flexible' halve can slightly move during closing, being engaged by the 'rigid' one and thus fixed to the axis of the VELO detector. All surfaces seen by the beam must be covered with a low Secondary Electron Emission (SEY) coating, like NEG or amorphous Carbon (a-C).

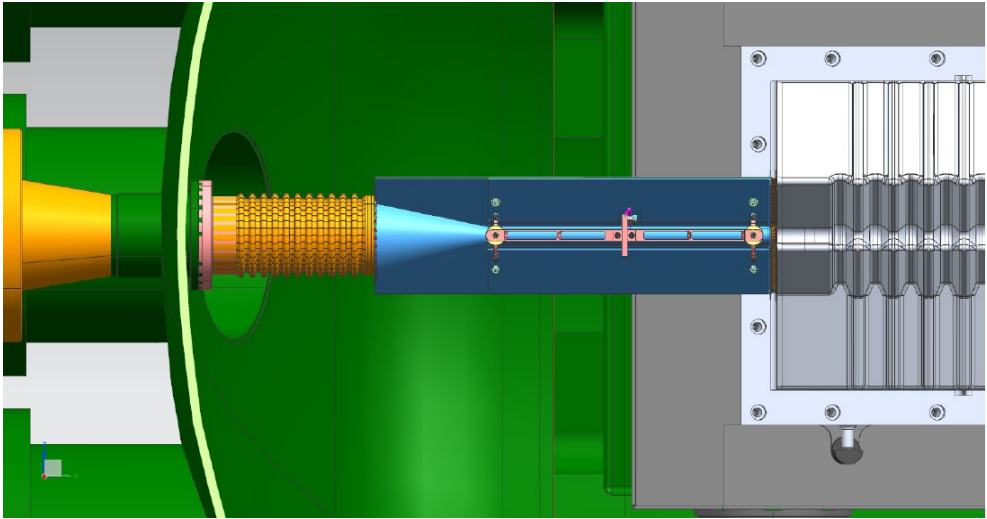

**Figure 3:** Side view of the SMOG2 storage cell system (see text). The beam employed for FT measurements is directed from left to right. The secondary particles move to the right into the VELO detector (light grey).

## 4. Polarized Gas Target for the LHC - Overview

At the LHC with its unpolarized beams, *Single Spin Transverse Asymmetries* (SSAA) can be measured with a transversely polarized target via the Φ-dependence of the final-state hadrons. SSTA's with electrons on a transversely polarized proton target have been studied at HERMES (DESY) [4].

For measuring the azimuthal distribution, an existing detector can be employed. Here, the most suitable LHC detector is the forward spectrometer of LHCb. It is proposed to install a polarized HERMES-like gas target internal to the LHC beam and within the acceptance of the LHCb detector. The main aim of the study group is at present on building and installing SMOG2 in order to take fixed target data in the next run(s).

In parallel to the SMOG2 project, a conceptual design is being studied, investigating critical machine issues which make the operation of a polarized gas target in the LHC beam impossible. Such items are:

(i)       a cell system of conductive surfaces with low SEY to prevent instabilities by the formation of electron clouds (EC);

(ii)      a cell surface with low recombination to avoid polarization losses, in accordance with the restrictions imposed by a safe operation of the LHC;

(iii)     a target chamber with transverse guide field, an openable cell system and an effective differential pumping of the target gas;

(iv)     optimum conditions for low Beam-Induced Depolarization (BID) of the target atoms by the periodic bunch fields.

**4.1 Suppressing of beam instabilities**

The cell system consisting of conductive surfaces may cause instabilities of two kinds: (a) by its shape which will determine its RF properties, e.g rapidly changing cross section or cavity-like structures, or (b) by surfaces with high yield of secondary electrons (high SEY) causing dense electron clouds trapped by the positive beam charge.

**(a)** By the bunch fields, excitation of excessive impedances in frequency space may occur, leading to energy dissipation in the surrounding chambers (beam induced heating), or to resonant kicks to the following bunches, resulting in beam blow up and losses. These effects can cause severe damage to the machine and its components, in particular at the LHC with its huge peak power where sophisticated collimator and beam dump systems are needed to protect the machine. All parts to be installed into the LHC are carefully inspected and approved by the Impedance group using numerical simulations and RF measurements. The SMOG2 target system shown in Fig. 3 has already been inspected [5] and approved as outlined in [3].

**(b)** The formation of dense Electron Clouds (EC) may lead to transverse oscillations of the beam with exponential growth, resulting in dangerous beam losses [6]. Evidently, the elevated residual gas pressure in the target region may favor EC formation, as well as chamber walls with high SEY. As coatings for surfaces close to the LHC beam, materials with SEY $\leq 1.4$ are allowed, only. These are (i) Non-Evaporable Getter (NEG) and (ii) amorphous Carbon (a-C).

NEG coating for the tube's inner surface is excluded because of its pumping action which will result in embrittlement and possible disintegration of the coating. At present, amorphous Carbon (a-C) seems to be the only viable solution. A-C is already applied in accelerators, including the SPS and the LHC. For coating beam screens of accelerator magnets, it can even be applied *in situ*, without removing the magnets.

For the coating of the polarized target cell, a-C is envisaged. The same coating will be applied for the SMOG2 cell (see Sec.3) which will yield during its operation valuable information for the design of the PGT. Clearly, an uncovered a-C surface exhibits strong recombination of the (polarized) atomic hydrogen atoms, reducing the target polarization to $\leq 0.5$, unmeasurable by the Breit-Rabi Polarimeter (BRP) [7]. At HERMES, the Al cell with Drifilm coating (forbidden at the LHC) was kept at $\leq 100$ K resulting in a formation of a thin ice layer which had ideal properties in respect to recombination and depolarization [7]. Such a mode of operation is envisaged for the LHC-PGT, too. Here, detailed studies on the SEY and the presence of thin frozen layers controlled by the cell temperature are required.

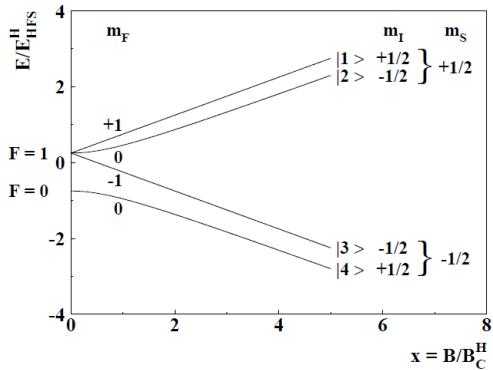

**Figure 4:** Hfs-diagram of hydrogen. The energy of the substates is shown as function of the external B-field in units of $B_c = 50.7$ mT. F = total spin; I, s = nuclear and electron spin. The substates are denoted by |1> - |4> with decreasing energy.

### 4.2 Beam-induced depolarization (BID)

Depolarization of the HERMES Polarized Gas Target by the HERA e± beams had been discussed in the 1989 proposal and provisions for high target polarization were taken in the course of the target development. The experimental study during with longitudinal guide field $B_=$ and optimization is described in Ref. [8]. Additional measurements with transverse $B_\perp$ together with the analysis of all measurements were subject of a dissertation by Phil Tait (Erlangen 2006) [9]. Its application to the LHC is discussed in a recent report [10].

BID is based on resonant transitions caused by the beam field acting on the polarized H-atoms in an external 'strong' guide field $B_0$ ($\approx 300$ mT). There are different classes of transitions, depending on the relative orientation of the guide field $\mathbf{B_0}$ and $\mathbf{B_1}$, the component of the beam field. With total spin F (for the notation see Fig.4 and Ref. 2) one has two classes of resonances:

- π resonances for $B_1 \perp B_0$ with selection rule $\Delta F = 0, \pm 1$ and $\Delta m_F = \pm 1$, and
- σ resonances for $B_1 \parallel B_0$ with selection rule $\Delta F = \pm 1$ and $\Delta m_F = 0$.

Some of these resonances change nuclear polarization. For longitudinal guide field, only the π resonances are present. For transverse field, like for HERMES run 2 or at LHCSpin, both types of resonances are present. The σ resonances, interchanging states 2 and 4, are densely spaced, i.e. its prevention requires a very high homogeneity of the guide field.

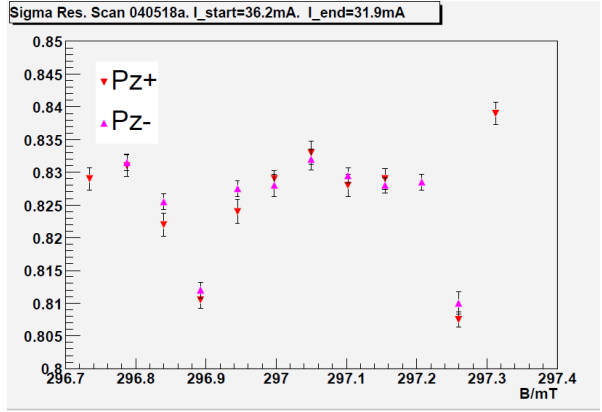

**Figure 5:** Two polarization minima caused by the $\sigma_{24}$ resonances are clearly visible demonstrating the high field quality of the HERMES target magnet. For the working point in between the resonances, a 2% increase in polarization is obtained (taken from [9]).

At HERMES during transverse running (2001 – 2004), careful studies to optimize the target polarization have been performed. As diagnostic tool, the target polarimeter with its capability of detecting single substates has been used, demonstrated in Fig.5. It shows the $\sigma_{24}$ resonances in a small B-range measured by the target polarimeter (BRP).

During the next LHC running period, a beam current of about 1 A is envisaged, which is 25 times higher than at HERA-e. A first attempt of a preliminary comparison is shown in a CERN-PBC note [10]. The relevant beam parameters of HERA-e and LHC are shown in the following **Table 2,** taken from [10].

The main differences between HERA-e and LHC are:

| Machine | $N_{Bunch}$ $x10^{10}$ | $f_{Bunch}$ MHz | $I_{Beam}$ A | $\sigma_z$ cm | $\sigma_t$ ps | $\alpha$ $ps^2$ | 1/e width F.S. | $I_0$ peak current(A) |
|---|---|---|---|---|---|---|---|---|
| **HERA-e** | 2.4 | 10.41 | 0.04 | 0.93 | 31 | $5.203 \cdot 10^{-4}$ | 5.1 GHz | 36.4 |
| **LHC** | 15.5 | 40.08 | 1.0 | 7.55 | 253 | $7.81 \cdot 10^{-6}$ | 0.63 GHz | 55.7 |

*Bunch frequency $f_B$:* 10.41 MHz vs. 40.08 MHz. This results for the LHC in a 4x larger spacing of the 'resonant B-values' (see Fig. 5), i.e. it lowers the requirements on field quality.

*Width of the Fourier spectrum:* 5.1 GHz vs. 0.63 GHz. This leads to a rapid fall-off of the relevant Fourier amplitudes of the $\sigma_{2-4}$ resonance (8.54 GHz) at the LHC (see Fig.5).

This naïve comparison suggests that BID at the LHC is probably less dangerous as at HERA-e, thanks to the higher bunch frequency and the much longer bunches, and despite the 25 times higher beam current. Clearly, this should be confirmed by a systematic study of BID at the LHC.

## 5.    Arrangement in the tunnel

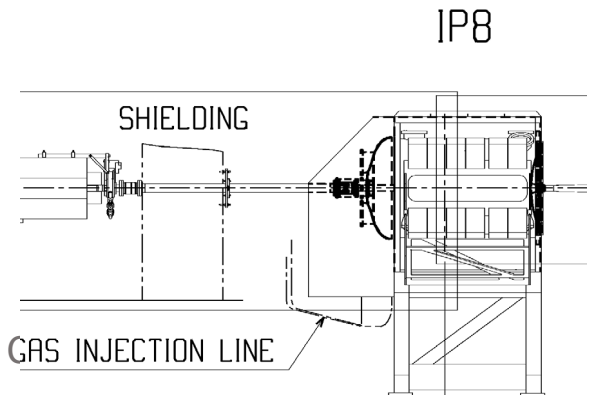

The PGT has to fit into the space upstream of the VELO vessel up to a shielding wall, which leaves about 1 meter for the target. Moving the wall would give more space, but it is important that the PGT stays in the acceptance of the VELO detector. The minimum distance to the VELO is given by considerations on effective differential pumping of the peaked flow from the downstream cell end.

In the horizontal transverse direction, there is enough space to place the Atomic Beam

Source (ABS) and diagnostics in the horizontal plane, which implies a vertical guide field **B₀**.

## 6. Vacuum system and estimate of the target density

The arrangement of the PGT in the beam line upstream of the VELO is shown schematically in Fig. 7. An intermediate configuration for Run 3 including a valve for sectoring the LHC vacuum is shown which allows for the installation of fixed-target experiments upstream of VELO without breaking the vacuum in the VELO vessel. The PGT cannot be located close to or even inside the VELO vessel because of the high gas flow of atomic and recombined hydrogen which requires differential pumping on a separate target chamber.

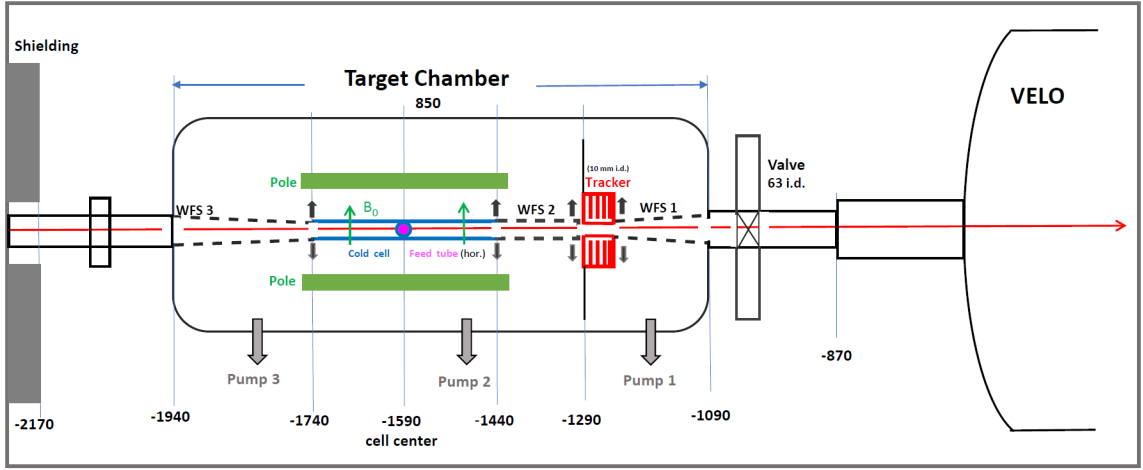

**Figure 7:** The conceptual drawing shows a sideview of the proposed PGT arrangement upstream of VELO with (right to left): new sectoring valve (installed in 2020), target chamber with conical WFS1 (L = 20 cm), Tracker (red) with 1 cm opening, gridded tube WFS 2 (L = 15 cm) for differential pumping, cold cell (L = 30 cm) with beam tube and feed tube for injection of ABS beam in a vertical guide field $B_0$ of about 0.3 T. The different pumping stages are indicated. The openable components are denoted with black arrows. The approximate positions are denoted by z (in mm). The IP8 is within the VELO detector at z = 0.

An estimate of the peaked flow out of the downstream cell opening shows that for the dimensions assumed, about 4 % of the down-stream flow passes the 1 cm opening in the Tracker, 30 cm downstream of the cell center. For the full ABS flow, this would correspond to about $10^{-6}$ mb l/s of recombined $H_2$. This is within the tolerable range.

The target density is estimated with the following assumptions:

$I_H$ (100 % HERMES ABS flow) = $6.5 \cdot 10^{16}$/s, may be limited by LHC vacuum constraints or space limitations for the PGT; cell 30 cm long, 1.0 cm inner diameter (i.d.), at 100K, with standard feed tube 10 cm long, 1.0 cm i.d.

The resulting 100% density of the polarized gas target is

θ = $1.2 \cdot 10^{14}$/cm², comparable with HERMES.

For the future HL-LHC-25ns, the maximum luminosity achievable with the PGT would be up to $8.3 \cdot 10^{32}$/cm² s. To which extent such densities can be realized and exploited in a real experiment, depends on many factors. This has to be investigated in more detail.

## 7. Outlook

As a first step towards a Polarized Gas Target for unpolarized gases in LS2 (2019/20), LHCSpin is seeking approval for SMOG2 in order to install in LS2 (2019/20). This will strongly improve the conditions for FT measurements by the LHCb experiment. In addition, it represents an important test of a storage cell at the LHC.

In parallel, the development of a PGT will continue. The key element, a target chamber with cooled cell to open, differential pumping and tracking detector, has highest priority. If built and tested during Run3 (2021-23), it could be installed in LS3 and tested using a gas feed system. Installation of the other PGT components during Technical Stops appears feasible.

The key parameters of the LHCSpin target are promising, as well as the broad physics potential, see [11]. With enough support from our Community it could become a success!

## Acknowledgement

The authors are indebted to P. Chiggiato, R. Cimino, M. Contalbrigo, R. Engels, M. Ferro-Luzzi, W. Funk, G. Graziani, K. Grigoryev, G. Iadarola, A. Nass, F. Rathmann, D. Reggiani, B. Salvant, and E. Thomas for discussions and valuable advice.

---

[2]Note added in proof: In December 2018, the SMOG2-EDR has been approved by the LHCb collaboration.